\documentclass[aps,prl,twocolumn,superscriptaddress,showpacs,preprintnumbers]{revtex4}
\usepackage{graphicx}

\begin{document}

\title{The Dichotomy between Nodal and Antinodal Quasiparticles in Underdoped (La$_{2-x}$Sr$_x$)CuO$_4$ Superconductors}

\author{X. J. Zhou$^{1,2}$, T. Yoshida$^{1,3}$, D.-H. Lee$^{4,5}$,  W. L. Yang$^{2,1}$,
V. Brouet$^{1,2}$, F. Zhou$^{6}$, W. X. Ti$^{6}$, J. W.
Xiong$^{6}$,  Z. X. Zhao$^{6}$, T. Sasagawa$^{1}$, T.
Kakeshita$^{7}$, H. Eisaki$^{7}$, S. Uchida$^{7}$, A.
Fujimori$^{3}$, Z. Hussain$^{2}$ and Z.-X. Shen$^{1}$ }

\affiliation{
\\$^{1}$Dept. of Physics, Applied Physics and
Stanford Synchrotron Radiation Laboratory, Stanford University,
Stanford, CA 94305
\\$^{2}$Advanced Light Source, Lawrence Berkeley National
Lab, Berkeley, CA 94720
\\$^{3}$Dept. of Physics, University of Tokyo, Bunkyo$-$ku, Tokyo 113, Japan
\\$^{4}$Dept. of Physics, University of California at Berkeley,
Berkeley, CA94720.
\\$^{5}$Material Science Division, Lawrence Berkeley National Lab, Berkeley, CA 94720
\\$^{6}$National Lab for Superconductivity, Institute of Physics,
Chinese Academy of Sciences, Beijing 100080, China
\\$^{7}$Dept. of Superconductivity, University of Tokyo,
Bunkyo$-$ku, Tokyo 113, Japan}

\date{June 26, 2003}

\begin{abstract}

High resolution angle-resolved photoemission measurements on
underdoped (La$_{2-x}$Sr$_x$)CuO$_4$ system show that,  at
energies below 70 meV, the quasiparticle peak is well defined
around the ($\pi$/2,$\pi$/2) nodal region and disappears rather
abruptly when the momentum is changed from the nodal point to the
($\pi$,0) antinodal point along the underlying ``Fermi surface''.
It indicates that there is an extra low energy scattering
mechanism acting upon the antinodal quasiparticles. We propose
that this mechanism is the scattering of quasiparticles across the
nearly parallel segments of the Fermi surface near the antinodes.

\end{abstract}

\pacs{74.25.Jb,71.18.+y,74.72.Dn,79.60.-i}

\maketitle

The high temperature superconductivity in cuprates is derived from
doping the parent antiferromagnetic Mott insulators.  It is found
that the normal state properties of cuprates are highly anomalous,
particularly in the underdoped region. Understanding the normal
state is believed to be a key for understanding the mechanism of
high temperature superconductivity\cite{Anderson}.

For the underdoped cuprates, one peculiar behavior of its
electronic structure, as revealed by angle-resolved photoemission
spectroscopy (ARPES), is the dichotomy between the
$\sim(\pi/2,\pi/2)$ nodal and $\sim(\pi,0)$ antinodal excitations.
In underdoped Bi$_2$Sr$_2$CaCu$_2$O$_8$ (Bi2212), it was found
that the lineshape of the {\it normal state} photoemission spectra
is broad in the antinodal region but sharp near the nodal
region\cite{ShenSchrieffer}.  %%In addition, while the nodal
%%excitation is insensitive to temperature, the antinodal spectra
%%change dramatically across the superconducting transition
%%temperature ($T_c$); below $T_c$ a sharp quasiparticle peak grows
%%out of the broad background.
It was proposed that the antinodal spectral broadening in the
normal state is due to the coupling of electrons with the
($\pi$,$\pi$) magnetic excitations\cite{ShenSchrieffer}. In the
supercnducting state, the antinodal spectrum is also believed to
be influenced by the ($\pi$,$\pi$) spin resonance
mode\cite{Campuzano}.
%%In addition, the antinodal
%%peak-dip-hump structure in the superconducting state has been
%%argued as a further evidence of the close relationship between the
%%($\pi$,$\pi$) spin resonance mode and
%%superconductivity\cite{Campuzano}.

The peculiar electronic structure of the underdoped sample may
provide important clues for understanding the anomalous normal
state properties.  It is therefore essential to establish whether
such a nodal-antinodal dichotomy is universal in cuprate
materials, and particularly, to establish its physical origin.
However, Bi2212 is not an ideal system for such an in-depth
investigation. Because of disorder, no sharp nodal structure has
been observed in deeply underdoped Bi2212. Furthermore, the states
near the antinodal region are severely complicated by its
superstructure. This, together with the bilayer splitting resolved
very recently\cite{FengBS}, raises concerns about the
interpretations\cite{JFink,MNorman,DessauKink}. The
La$_{2-x}$Sr$_x$CuO$_4$ (LSCO) system, in comparison, is an ideal
candidate to address the issue in the underdoped region because
the system becomes less disordered with decreasing doping. As we
will show, one can see an essentially resolution-limited sharp
nodal peak in LSCO with a doping level as low as 6.3$\%$ which has
not been observed in Bi2212 with a comparable doping. Its simple
crystal structure also makes it free from the complications of the
superstructure and bilayer splitting encountered in Bi2212.

While the majority of photoemission work so far has been performed
on Bi2212, data on LSCO are available only recently because of the
improved sample quality and better understanding of matrix element
effects involved in measuring LSCO
system\cite{ZhouPRL,YoshidaMatrix,ZhouNature}.  In this paper, we
present detailed angle-resolved photoemission data on underdoped
LSCO superconductors. We observe remarkably sharp nodal
quasiparticle peak at all doping levels studied, even for heavily
underdoped samples. In contrast, the antinodal peak only exists in
optimally-doped and overdoped samples. Particularly, these sharp
peaks can  be observed only at low energy, below 70 meV.  In
addition, for underdoped samples, when moving from nodal to
antinodal regions, we find that the disappearance of sharp peaks
occurs in a fairly abrupt fashion near where the Fermi surface
changes from parallel to the ($\pi$,0)-(0,$\pi$) diagonal to
parallel to the (0,0)-($\pi$,0) or (0,0)-(0,$\pi$) directions.
Intrigued by the close tie between the quasiparticle scattering
and the Fermi surface topology, we propose this ``nodal-antinodal
dichotomy'' is due to the scattering of quasiparticles across the
almost parallel segments of the Fermi surface near the antinodes.

The photoemission measurements were carried out on beamline 10.0.1
at Advanced Light Source, using a Scienta 2002 electron energy
analyzer. The photon energy is 55 eV and the $\vec{E}$-vector of
the incident light is parallel to the CuO$_2$ plane and 45$^\circ$
with respect to the Cu-O bond, as indicated by the arrow in Fig.
2a\cite{ZhouPRL,YoshidaMatrix}. The energy resolution is
15$\sim$20 meV and the angular resolution is 0.3$^\circ$
(corresponding to 0.018 $\AA$$^{-1}$ in momentum). In this paper
we mainly present our data on the underdoped LSCO (x=0.063,
T$_c$=12 K) and LSCO (x=0.09, T$_c$=28K) samples. For comparison,
we also show data on overdoped LSCO (x=0.22, T$_c$=24K). The LSCO
single crystals are grown by travelling solvent floating zone
method\cite{FZhou}. The samples were cleaved {\it in situ} in
vacuum with a base pressure better than 4$\times$10$^{-11}$ Torr
and measured at a temperature of $\sim$20K.

\begin{figure}[tbp]
\begin{center}
\includegraphics[width=0.65\columnwidth,angle=-90]{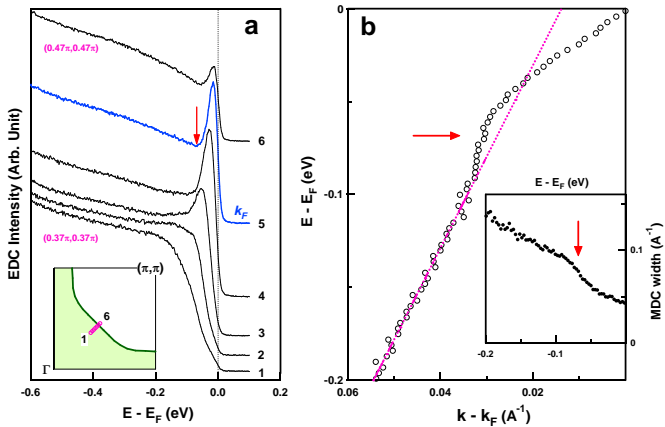}
\end{center}
\caption{Electronic structure of LSCO (x=0.063, T$_c$=12 K) along
the (0,0)-($\pi$,$\pi$) nodal direction (inset of Fig. 1a) at a
temperature of 20 K.  (a). EDCs measured along the nodal direction
in a second Brillouin zone. The EDCs are numbered according to the
momentum points in the inset. These momentum points are equal
spaced: point 1 corresponds to (0.374$\pi$,0.374$\pi$) and  point
6 corresponds to (0.466$\pi$,0.466$\pi$). The red arrow indicates
an energy of $\sim$70 meV below which the quasiparticle survives
and above which it turns into a broad edge. (b). A kink in the
dispersion at $\sim$70 meV as indicated by an arrow. The dotted
pink line is a guide to the eye which is a line fitting the
high-energy part of the dispersion. In the inset is the MDC width
which shows a drop at an energy of $\sim$70 meV, as indicated by
an arrow.}
\end{figure}

Fig. 1 presents experimental results along the (0,0)-($\pi$,$\pi$)
nodal direction of the LSCO (x=0.063) sample. Even for this
extremely underdoped sample in the vicinity of an
insulator-superconductor transition, one can see a remarkably
sharp quasiparticle peak in the nodal region with a clear
dispersion (Fig. 1a). The sharp peak abruptly turns into an edge
once it disperses up to an energy of $\sim$ 70meV.  Such a
dramatic change in spectral shape is not observed in underdoped
Bi2212, presumably due to stronger disorder in Bi2212.  The
dispersion (Fig. 1b), extracted by fitting momentum distribution
curves
(MDCs)\cite{PashaEnergy,KaminskiEnergy,PJohnsonEnergy,LanzaraEnergy,ZhouNature},
shows a clear slope break (a kink) at an energy $\sim$ 70meV. The
MDC width, which is related to the scattering rate $\tau^{-1}$,
also exhibits a drop at the same energy (inset of Fig. 1b). All
these observations indicate that there is an energy scale at
$\sim$70 meV for the nodal quasiparticles.

\begin{figure*}[floatfix]
\begin{center}
\includegraphics[width=0.72\columnwidth,angle=-90]{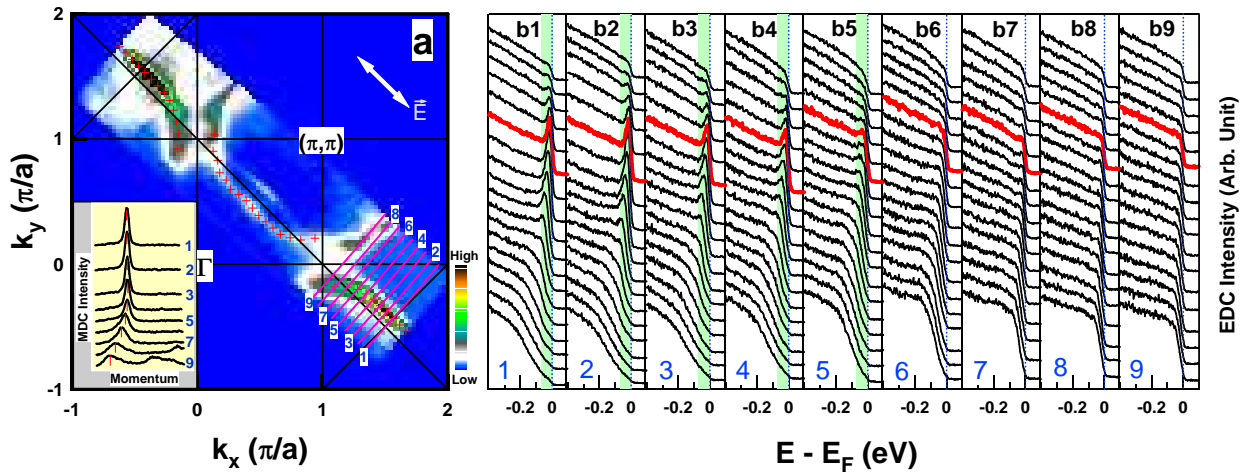}
\end{center}
\caption{(a). Low energy spectral weight as a function of k$_x$
and k$_y$ for the LSCO x=0.063 sample measured at a temperature of
20 K.  The inset shows MDCs at E$_F$ along several momentum cuts
in one octant of the second Brillouin zone; the corresponding cuts
are marked in the figure with a number from 1 to 9. The red cross
in the figure represents the peak position of MDCs at E$_F$.
(b1-b9). EDCs along the cuts as marked in (a). The red spectra are
EDCs  on the underlying Fermi surface. The green shades in (b1-b5)
highlight the energy range (0$\sim$70meV) within which sharp peaks
are confined.}
\end{figure*}

Fig. 2a shows the low-energy spectral weight of the LSCO (x=0.063)
sample as a function of momentum by integrating EDCs in a narrow
energy window near E$_F$ (-5meV,5meV) (k$_x$ and k$_y$ are along
Cu-O bonding direction). The high intensity contour constitutes
what we call the ``Fermi surface''\cite{NoteFS}. To be
quantitative, we used MDCs at E$_F$ to extract the Fermi momentum
(k$_F$) by following the MDC peak position. This is exemplified in
the inset of Fig. 2a for nine cuts in a second zone and the
obtained k$_F$s are marked as red crosses in Fig. 2a. The k$_F$s
are obtained in another second zone and the first zone in a
similar manner although the relative spectral intensity in the
first Brillouin zone is much weaker. Covering multiple Brillouin
zones allows us to align the sample to high precision because it
provides an internal check: the Fermi surfaces obtained from
different zones have to be consistent with each other. The final
extracted Fermi surface is plotted in Fig. 3. To a good
approximation, the measured Fermi surface can be represented as
three straight segments: two antinodal ones (marked black in Fig.
3) running parallel to (0,0)-($\pi$,0) and (0,0)-(0,$\pi$)
directions, respectively,  and the nodal one (marked red in Fig.
3) running parallel to ($\pi$,0)-(0,$\pi$) diagonal direction.

\begin{figure}[tbp]
\begin{center}
\includegraphics[width=0.82\linewidth,angle=0]{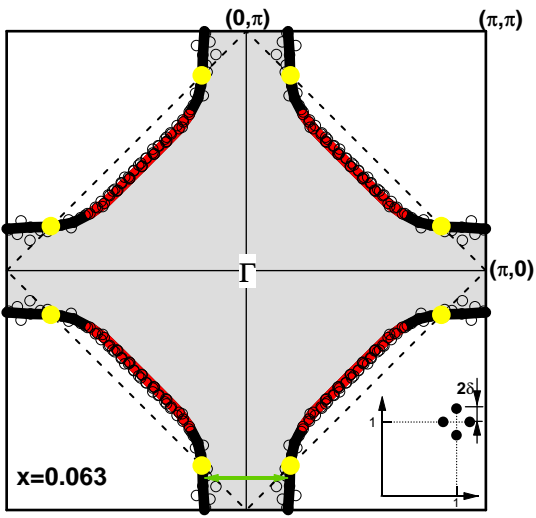}
\end{center}
\caption{Experimental Fermi surface for LSCO x=0.063 sample. The
black open circles are obtained from the MDC peak position at
E$_F$, as shown in Fig. 1 as the red crosses, and then symmetrized
in the first Brillouin zone. The solid lines are guides to the eye
for the measured Fermi surface. The red lines represent the
portion of Fermi surface where one can see quasiparticle peaks.
The dotted black line represents the antiferromagnetic Brillouin
zone boundary; its intersection with the Fermi surface gives eight
``hot spots" (solid yellow circles) from ($\pi$,$\pi$) magnetic
excitations. The double-arrow-ended green line represents a
nesting vector, (0.35$\pi$,0), between the antinodal part of the
Fermi surface. In the inset shows the schematic neutron
diffraction pattern observed in LSCO superconductors with four
incommensurate peaks of distance 2$\delta$ from ($\pi$,$\pi$)
point.}
\end{figure}

Fig. 2(b1-b9) shows the energy distribution curves (EDCs) along
nine cuts of the Fermi surface from the nodal to the antinodal
region (as marked in Fig. 2a) for the x=0.063 sample. The red
curves are the EDCs at $k=k_F$. It is clear that all quasiparticle
peaks are confined within 70 meV energy range near the Fermi
level. Moreover, the quasiparticle peaks exist only on part of the
Fermi surface near the nodes,  as marked by the solid red line in
Fig. 3.  Away from the ``nodal segment'' the peak gets weaker and
disappears in a fairly abrupt fashion. This can be seen more
clearly from Fig. 4a where the EDCs on the underlying Fermi
surface and at ($\pi$,0) are plotted together.  Similar sharp
transition is also observed in another underdoped LSCO (x=0.09)
sample (Fig. 4b). The situation for these underdoped samples is
very different from that in the highly overdoped (x=0.22) sample
(Fig. 4c). In that case we see quasiparticle peak over the entire
Fermi surface; the antinodal peak appears even sharper than the
nodal peak. This doping dependence clearly indicates that the
spectral broadening near the antinodal region in the underdoped
samples {\it is not due to a matrix element effect}. This is also
consistent with earlier observations in
Bi2212\cite{ShenSchrieffer}.

At first glance, the concept of quasiparticle seems entirely
inappropriate for heavily underdoped cuprates. Given the fact that
we are dealing with a strongly correlated system, the existence of
sharp nodal quasiparticle below 70 meV is itself a miracle, not to
mention the nodal-antinodal dichotomy. One might argue that the
nodal-antinodal dichotomy is due to the much higher excitation
energy near the antinodes compared to that near the nodes, as
often assumed in the theory literature. We stress that the
antinodal edge ($\sim$15 meV) discussed in this paper is not
particularly high in energy compared to that of some nodal peaks
(Fig. 4) and is definitely below 70 meV along the Fermi surface
locus. Therefore, this trivial explanation does not work.  The
spirit of our paper is to assume the existence of a mechanism that
allows {\it all} low energy quasiparticles below 70 meV.  Under
the working of this mechanism both low energy nodal and antinodal
excitations become sharp Bogoliubov-Landau quasiparticles.  The
Fermi surface map should also be taken in this context. Here we
put the phrase ``Fermi surface" in quotation marks to reflect the
fact that by that we mean the locus of the lowest energy
quasiparticle excitations before the extra scattering on antinodal
excitations is switched on. Then we ask what extra is needed to
explain the antinodal spectral broadening.

\begin{figure}[b]
\begin{center}
\includegraphics[width=0.68\columnwidth,angle=-90]{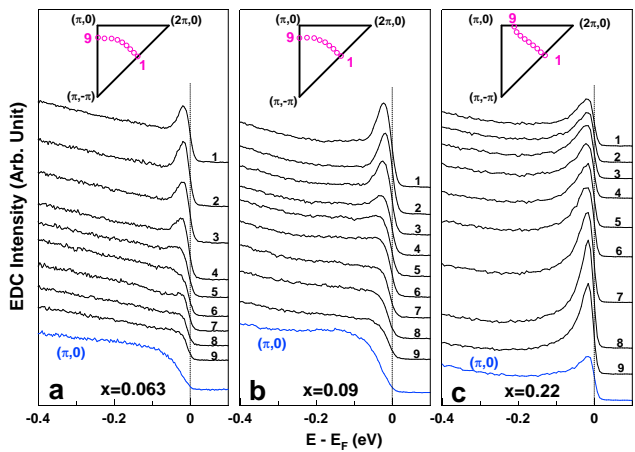}
\end{center}
\caption{EDCs on Fermi surface for LSCO x=0.063(a), 0.09(b) and
0.22(c) samples. All samples are measured at $\sim$20 K. The
corresponding momentum position is marked in the upper inset of
each panel. Also included are the spectra at ($\pi$,0) points,
colored as blue.}
\end{figure}

One candidate that immediately comes to mind is the ($\pi$,$\pi$)
magnetic fluctuation that produces ``hot spots" on the Fermi
surface, as previously proposed for Bi2212\cite{ShenSchrieffer}.
We note that neutron scattering has revealed a significant
difference in the magnetic response of LSCO and Bi2212. The
$(\pi,\pi)$ spin resonance mode observed in Bi2212\cite{BKeimer}
is absent in LSCO. Instead, incommensurate magnetic peaks were
observed at low energy (below 15
meV)\cite{Aeppli,Tranquada,KYamada} (inset of Fig. 3), which
broaden rapidly with increasing energy although the magnetic
fluctuation can persist up to 280 meV\cite{Hayden}. To check
whether the low energy incommensurate magnetic fluctuation can be
responsible for the lack of spectral peaks in the antinodal
segments, we have performed the following analysis. First, we
shift the measured Fermi surface by the peak wave vectors of the
magnetic excitation ($(\pi,\pi)\pm (2\delta,0)$ and $(\pi,\pi)\pm
(0,2\delta)$ with $\delta$ being the incommensurability) to
produce four Fermi surface replicas. Then we record the
intersections of these replicas with the original Fermi surface.
These intersections are ``hot spots" where the quasiparticles will
experience scattering from the incommensurate magnetic
fluctuations. For $x=0.063$ and $x=0.09$ samples the obtained hot
spots do concentrate around the antinodes. However, for $x=0.15$
and $x=0.22$ samples the above construction also leads to ``hot
spots" mainly near the antinodal segments but the quasiparticle
peak can be seen over the entire Fermi surface. Considering that
the incommensurate peaks are present in LSCO up to
x=0.25\cite{KYamada}, this latter observation is inconsistent with
the mechanism of the incommensurate magnetic fluctuations although
one can not completely rule out this possibility because how the
coupling strength varies with doping is not known.

Intrigued by the fact that the extra broadening sets in when the
Fermi surface turns from the ($\pi$,0)-(0,$\pi$) diagonal
direction to (0,0)-($\pi$,0) or (0,0)-(0,$\pi$) direction,  we
propose an alternative mechanism that the scattering in question
causes a pair of electrons on two parallel antinodal segments to
be scattered to the opposite ones (Fig. 3), {\it i.e.}, {\bf
p$_1$} =($\pm$0.175$\pi$,p$_{1y}$), {\bf
p$_2$}=($\mp$0.175$\pi$,p$_{2y}$) $\rightarrow$ {\bf
p$_1'$}=($\mp$ 0.175$\pi$, p$_{1y}$), {\bf p$_2'$}=($\pm$
0.175$\pi$, p$_{2y}$), or {\bf p$_1$}=(p$_{1x}$,$\pm$0.175$\pi$),
{\bf p$_2$}=(p$_{2x}$,$\mp$0.175$\pi$) $\rightarrow$ {\bf
p$_1'$}=(p$_{1x}$,$\mp$0.175$\pi$), {\bf p$_2'$}=(p$_{2x}$,$\pm$
0.175$\pi$).
%%\begin{eqnarray}
%%{\bf p_1}=(\pm0.175\pi,p_{1y}), {\bf p_2}=(\mp0.175\pi,p_{2y})
%%\rightarrow {\bf p_1'}=(\mp 0.175\pi, p_{1y}), {\bf p_2'}=(\pm
%%0.175\pi, p_{2y}),
%%\end{eqnarray}
%%or
%%\begin{eqnarray}
%%{\bf p_1}=(p_{1x},\pm0.175\pi), {\bf p_2}=(p_{2x},\mp0.175\pi)
%%\rightarrow {\bf p_1'}=(p_{1x},\mp0.175\pi), {\bf
%%p_2'}=(p_{2x},\pm 0.175\pi).
%%\end{eqnarray}
In the normal state this scattering mechanism can cause a
quasiparticle to decay into two quasiparticles and one quasihole.
The antinodal spectral broadening occurs as a result of the
frequent occurrence of such a decay which renders the normal state
quasiparticle ill-defined.

In summary, we have shown that the low energy excitations between
nodal and antinodal quasiparticles behave very differently in the
underdoped superconductors. Evidently such a dichotomy is due to
the existence of an extra {\it low energy} scattering mechanism
that operates primarily on antinodal quasiparticles. We propose
this may be associated with quasiparticle scattering across the
nearly parallel segments of the Fermi surface near the antinodes.
Clearly this proposal requires more scrutiny, both experimentally
and theoretically.


\begin{thebibliography}{99}
\bibitem{NoteFS} Throughout the paper we use the phrase
``Fermi surface'' loosely to denote the momentum space locus of
high intensity low-energy spectral weight.
\end{thebibliography}
\end{document}